\documentclass[prl,superscriptaddress,preprintnumbers,twocolumn]{revtex4}
\usepackage[ansinew]{inputenc}
\usepackage{graphicx}
\usepackage{amsmath}
\usepackage{amsthm}
\usepackage{bm}
\usepackage{layout}
\usepackage{float}
\usepackage{amsfonts}
\usepackage{amssymb}

\usepackage[pdftex,
colorlinks=true,
linkcolor=blue,citecolor=blue,urlcolor=blue,]{hyperref} 
\begin{document}
\preprint{Accepted for the 2008 IAC Microgravity Sciences and
Processes Symposium, Sept.\ 29 to Oct.\ 3, in Glasgow, Scotland,
UK\vspace{1cm}}

\title{\huge{Space-QUEST}\\ \vspace{0.1cm} \Large{Experiments with quantum entanglement in space}}

\author{ \vspace{0.125cm}   Rupert  Ursin   }
\affiliation{ Faculty of Physics, University of Vienna, Austria
}\email{Rupert.Ursin@univie.ac.at}
\homepage{http://www.quantum.at/quest} \affiliation{   Institute for
Quantum Optics and Quantum Information (IQOQI), Austrian Academy of
Sciences, Austria }
\author{    Thomas  Jennewein   }   \affiliation{   Institute for Quantum Optics and Quantum Information (IQOQI), Austrian Academy of Sciences, Austria }
\author{    Johannes    Kofler  }   \affiliation{   Faculty of Physics, University of Vienna, Austria   }
                \affiliation{   Institute for Quantum Optics and Quantum Information (IQOQI), Austrian Academy of Sciences, Austria }
\author{    Josep M.    Perdigues   }   \affiliation{   European Space Research and Technology Centre (ESTEC), European Space Agency (ESA), The Netherlands }
\author{    Luigi   Cacciapuoti }   \affiliation{   European Space Research and Technology Centre (ESTEC), European Space Agency (ESA), The Netherlands }
\author{    Clovis J.   de Matos    }   \affiliation{   European Space Research and Technology Centre (ESTEC), European Space Agency (ESA), The Netherlands }
\author{    Markus  Aspelmeyer  }   \affiliation{   Institute for Quantum Optics and Quantum Information (IQOQI), Austrian Academy of Sciences, Austria }
\author{    Alejandra   Valencia    }   \affiliation{   Institute of Photonic Sciences (ICFO), Spain    }
\author{    Thomas  Scheidl }   \affiliation{   Institute for Quantum Optics and Quantum Information (IQOQI), Austrian Academy of Sciences, Austria }
\author{    Alessandro Fedrizzi }   \affiliation{   Institute for Quantum Optics and Quantum Information (IQOQI), Austrian Academy of Sciences, Austria }
\author{    Antonio Acin    }   \affiliation{   Institute of Photonic Sciences (ICFO), Spain    }
\author{    Cesare  Barbieri    }   \affiliation{   Dipartimento di Astronomia, University of Padova, Italy }
\author{    Giuseppe    Bianco  }   \affiliation{   Matera Space Geodesy Center, Agenzia Spaziale Italiana (ASI), Italy }
\author{    Caslav  Brukner }   \affiliation{   Faculty of Physics, University of Vienna, Austria   }
                \affiliation{   Institute for Quantum Optics and Quantum Information (IQOQI), Austrian Academy of Sciences, Austria }
\author{    Jos\'e  Capmany }   \affiliation{   Institute of Telecommunications and Multimedia Applications (ITEAM), Universidad Politécnica de Valencia, Spain }
\author{    Sergio  Cova    }   \affiliation{   Dip.\ Elettronica e Informazione, Politecnico di Milano, Italy  }
\author{    Dirk    Giggenbach  }   \affiliation{   Institut f\"ur Kommunikation und Navigation, Deutsches Zentrum f\"ur Luft- und Raumfahrt (DLR), Oberpfaffenhofen, Germany   }
\author{    Walter  Leeb    }   \affiliation{   Institute of Communications and Radio-Frequency Engineering, Vienna University of Technology, Austria   }
\author{    Robert H.   Hadfield    }   \affiliation{   School of Engineering\ and Physical Sciences, Heriot-Watt University, UK    }
\author{    Raymond     Laflamme    }   \affiliation{   Department of Physics \& Astronomy, Institute for Quantum Computing, University of Waterloo, Canada }
\author{    Norbert L\"utkenhaus    }   \affiliation{   Department of Physics \& Astronomy, Institute for Quantum Computing, University of Waterloo, Canada }
\author{    Gerard  Milburn }   \affiliation{   Department of Physics, University of Queensland, Australia  }
\author{    Momtchil    Peev    }   \affiliation{   Smart Systems, ARC - Austrian Research Centers GmbH, Austria    }
\author{    Timothy Ralph   }   \affiliation{   Department of Physics, University of Queensland, Australia  }
\author{    John    Rarity  }   \affiliation{   Department of Electrical and Electronic Engineering, University of Bristol, UK  }
\author{    Renato  Renner  }   \affiliation{   Swiss Federal Institute of Technology (ETH Z\"urich), Switzerland   }
\author{    Etienne Samain  }   \affiliation{   Observatoire de la Cote d'Azur, France  }
\author{    Nikolaos    Solomos }   \affiliation{   Pure and Applied Physics Laboratories, Hellenic Naval Academy, Piraeus, Greece  }
                \affiliation{   EUDOXOS Observatory, Kefalinia, Greece  }
\author{    Wolfgang    Tittel  }   \affiliation{   Institute for Quantum Information Science, University of Calgary, Canada    }
\author{    Juan P.     Torres  }   \affiliation{   Institute of Photonic Sciences (ICFO), Spain    }
\author{    Morio   Toyoshima   }   \affiliation{   Space Communication Group, National Institute of Information and Communications Technology (NICT), Japan    }
\author{    Arturo  Ortigosa-Blanch }   \affiliation{   Institute of Telecommunications and Multimedia Applications (ITEAM), Universidad Politécnica de Valencia, Spain }
\author{    Valerio Pruneri }   \affiliation{   Institute of Photonic Sciences (ICFO), Spain    }
                \affiliation{   Instituci\'{o} Catalana de Recerca i Estudis Avançats (ICREA), Spain    }
\author{    Paolo   Villoresi   }   \affiliation{   Department of Information Engineering (DEI), University of Padova, Italy    }
                \affiliation{   Istituto Nazionale per la Fisica della Materia - Consiglio Nazionale delle Ricerche (INFM-CNR), Italy   }
\author{    Ian Walmsley    }   \affiliation{   Atomic and Laser Physics, Clarendon Laboratory, University of Oxford, UK    }
\author{    Gregor  Weihs   }   \affiliation{   Department of Physics \& Astronomy, Institute for Quantum Computing, University of Waterloo, Canada }
\author{    Harald  Weinfurter  }   \affiliation{   Department f\"ur Physik, Ludwig-Maximilians-Universit\"at (LMU),  Munich, Germany   }
\author{    Marek   \.{Z}ukowski    }   \affiliation{   Institute for Theoretical Physics and Astrophysics, University of Gdansk, Poland    }
\author{    Anton   Zeilinger   }   \affiliation{   Faculty of Physics, University of Vienna, Austria   }
                \affiliation{   Institute for Quantum Optics and Quantum Information (IQOQI), Austrian Academy of Sciences, Austria }


\date{\today}

\begin{abstract}
\vspace{0.1cm}\textbf{The European Space Agency (ESA) has supported
a range of studies in the field of quantum physics and quantum
information science in space for several years, and consequently we
have submitted the mission proposal Space-QUEST (\textbf{Qu}antum
\textbf{E}ntanglement for \textbf{S}pace Experimen\textbf{t}s) to
the European Life and Physical Sciences in Space Program. We propose
to perform space-to-ground quantum communication tests from the
International Space Station (ISS). We present the proposed
experiments in space as well as the design of a space based quantum
communication payload.}
\end{abstract}

\maketitle


\section{Scientific Background}
Quantum entanglement is, according to Erwin Schr\"odinger in 1935
\cite{Schrodinger35a}, the essence of quantum physics and inspires
fundamental questions about the principles of nature. By testing the
entanglement of particles we are able to ask fundamental questions
about realism and locality in nature \cite{Bell64a,Leggett03}. Local
realism imposes certain constraints in statistical correlations of
measurements on multi-particle systems. Quantum mechanics, however,
predicts that entangled systems have much stronger than classical
correlations that are independent of the distance between the
particles and are not explicable with classical physics.

It is an open issue whether quantum laws, originally established to
describe nature at the microscopic level of atoms, are also valid in
the macroscopic domain such as long distances. Various proposals
predict that quantum entanglement is limited to certain mass and
length scales \cite{Ghirardi86a,Penrose96} or altered under specific
gravitational circumstances \cite{Ralph06}.

Testing the quantum correlations over distances achievable with
systems placed in the Earth orbit or even beyond
\cite{Kaltenbaek03} would allow to verify both the validity of
quantum physics and the preservation of entanglement over distances
impossible to achieve on ground.

Using the large relative velocity of two orbiting satellites, one
can perform experiments on entanglement where---due to special
relativity---\textit{both} observers can claim that they have
performed the measurement on their system prior to the measurement
of the other observer. In such an experiment it is not possible
anymore to think of any local realistic mechanisms that potentially
influence one measurement outcome according to the other one.

Moreover, quantum mechanics is also the basis for emerging
technologies of quantum information science, presently one of the
most active research fields in physics. Today's most prominent
application is quantum key distribution (QKD) \cite{Gisin02a}, i.e.\
the generation of a provably unconditionally secure key at distance,
which is not possible with classical cryptography. The use of
satellites allows for demonstrations of quantum communication on a
global scale, a task impossible on ground with current optical fiber
and photon-detector technology. Currently, quantum communication on
ground is limited to the order of 100 of kilometers
\cite{Waks02,Takesue05}. Bringing quantum communication into space
is the only way to overcome this limit with state-of-the-art
technology.

Another area of applications is in metrology, where quantum clock
synchronization and  quantum positioning \cite{Valencia02} are
studied. Furthermore, sources of quantum states in space may have
applications in the new field of quantum astronomy \cite{Naletto07}.

\section{The Proposal} We propose to perform these experiments in
space by placing a quantum transceiver on the external pallet of the
European Columbus module at the ISS. The entire terminal must not
exceed the specifications given for pallet payloads as provided by
ESA \cite{Carey01}. The requirements are: size
$1.39×1.17×0.86$~m$^3$, mass $<100$~kg, and a peak power consumption
of $<250$~W, respectively. A preliminary design of a satellite-based
quantum transceiver (including an entangled photon source, a weak
pulse laser sources, single photon detection modules together with
two transceiver telescopes) based on state-of-the-art optical
communication terminals and adapted to the needs of quantum
communication is already published in \cite{Pfennigbauer05}.

The entangled photons are transmitted to two distant ground stations
via simultaneous down-links \cite{aspelmeyer03a}, allowing a test on
entanglement and the generation of an unconditional secure quantum
cryptographic key between stations separated by more than 1000~km.

Additionally, such a quantum transceiver in space is capable of
performing two consecutive single down-links---using the entangled
or the weak pulse laser onboard the satellite---establishing two
\textit{different} secure keys between the satellite and each of the
ground stations (say, Vienna and Tokyo). Then a logical combination
of the two keys (e.g.\ bitwise XOR) is sent publicly to one of the
two ground stations. Out of that \textit{one} unconditionally secure
key \textit{between} the two ground stations can be computed. Using
such a scheme would allow for the first demonstration of global
quantum key distribution. Furthermore an uplink scenario is
published in \cite{Rarity02}.

An important step towards the applicability of quantum communication
on a global scale, is to extend single QKD links to a quantum
network by key relaying along a chain of trusted nodes
\cite{Poppe08,Dianeti08} using satellites as well as fiber-based
systems. Furthermore, the efficiency of quantum networks can be
improved employing quantum percolation protocols \cite{Acin07}.

It would be favorable to include in parallel to the QKD down-link
from the ISS a high-speed communication link providing several
Gigabit per second bandwidth \cite{Toyoshima07b,Giggenbach07}.

\begin{figure}[t]\label{quest}
\includegraphics[width=7cm]{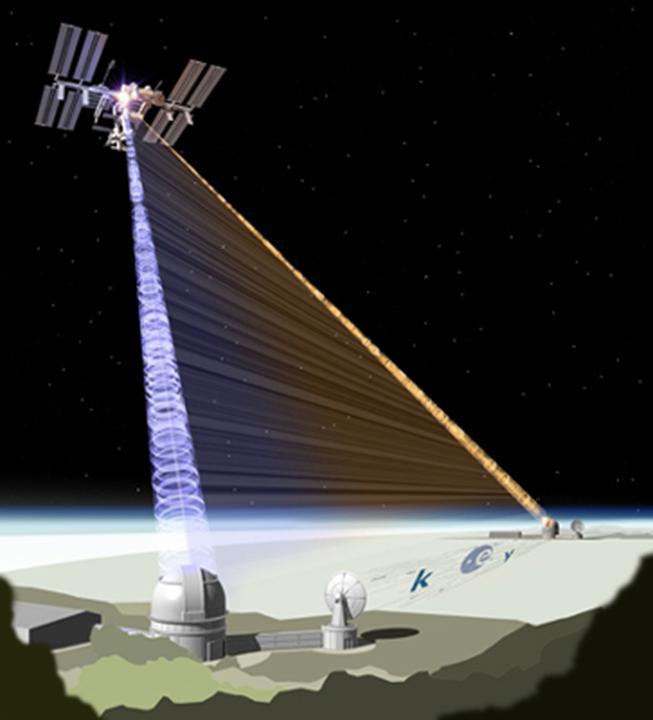}
\caption{Distribution of pairs of entangled photons using the
International Space Station (ISS). Entangled photon pairs are
simultaneously distributed to two separated locations on Earth, thus
enabling both fundamental quantum physics experiments and novel
applications such as quantum key distribution.}
\end{figure}

\section{Proof-of-principle experiments} As an important step
towards  quantum communication protocols using satellites various
proof-of-principle demonstrations of quantum communication protocols
have already been performed over terrestrial free-space links
\cite{hughes02,kurtsiefer02,Aspel03c,pan05a}. One experiment was
carried out on the Canary islands using a 144 km free-space link,
between the neighboring islands La Palma and Tenerife (Spain), where
ESA's 1-meter-diameter receiver telescope, originally designed for
classical laser communication with satellites, was used
\cite{Schmitt07,ursin07} to receive single photons.

In a second experiment the Matera-Laser-Ranging-Observatory (Italy)
was used to establish a single photon down-link from a low-earth
orbit satellite \cite{Villoresi08}. A satellite-to-Earth
quantum-channel down-link was simulated by reflecting attenuated
laser pulses off the optical retroreflector on board of the
satellite Ajisai, whose orbit has a perigee height of 1485~km.

An important component in space based quantum communication is a
source for entangled photons, that is suitable for space
applications in terms of efficiency, mass and power consumption. A
source fulfilling the payload requirements based on highly efficient
down-conversion crystals which deliver the necessary numbers of
photon pairs is published in \cite{Fedrizzi07}.

\section{Topical Team}
In 2007 the formation of a Topical Team for supporting the
Space-QUEST experiment comprised of researchers from academia
actively involved in relevant scientific fields was initiated by ESA
and currently consists of 27 members from 10 countries. This team
will support the proposal with their individual scientific and
technical expertise and also aims to increase the research
community's interaction with industry. The present programmatic
roadmap of Space-QUEST is compatible with a launch date by end of
2014 \cite{Perdigues07}.

\section{Conclusions}
We emphasize that the space environment will allow quantum physics
experiments with photonic entanglement and single photon quantum
states to be performed on a large, even global, scale. The
Space-QUEST proposal aims to place a quantum communication
transceiver containing the entangled photon source, a weak pulsed
(decoy) laser source and single photon counting modules in space and
will accomplish the first-ever demonstration in space of fundamental
tests on quantum physics and quantum-based telecom applications. The
unique features of space offer extremely long propagation paths to
explore the limits of the validity of quantum physic's principles.
In particular, this system will allow for a test of quantum
entanglement over a distance exceeding 1000~km, which is impossible
on ground.

\section{Acknowledgments} This work was supported by the European
Space Agency under contract numbers 16358/02/NL/SFe, 17766/03/NL/PM
and 18805/04/NL/HE as well as by the national space delegations.
Additional funding was provided by the European Commission (QAP).


\begin{thebibliography}{10}

\bibitem{Schrodinger35a}
E.~Schr{\"o}dinger.
\newblock Die gegenw{\"a}rtige {S}ituation in der {Q}uantenmechanik.
\newblock {\em Naturwissenschaften}, 23:807--812; 823--828; 844--849, 1935.

\bibitem{Bell64a}
J.~S. Bell.
\newblock On the {E}instein {P}odolsky {R}osen paradox.
\newblock {\em Physics}, 1:195--200, 1964.

\bibitem{Leggett03}
A.~J. Leggett.
\newblock Nonlocal hidden-variable theories and quantum mechanics: An
  incompatibility theorem.
\newblock {\em Found. Phys.}, 33:1469–1493, 2003.

\bibitem{Ghirardi86a}
G.~C. Ghirardi, A.~Rimini, and T.~T.~Weber.
\newblock Unified dynamics for microscopic and macroscopic systems.
\newblock {\em Phys. Rev. D}, 34:470, 1986.

\bibitem{Penrose96}
R.~Penrose.
\newblock On gravity's role in quantum state reduction.
\newblock {\em Gen. Rel. Grav.}, 28:581, 1996.

\bibitem{Ralph06}
T.~C. Ralph, G.~J. Milburn, and T.~Downes.
\newblock Gravitationally induced decoherence of optical entanglement.
\newblock {\em arXiv:quant-ph/0610093v1}.

\bibitem{Kaltenbaek03}
R.~Kaltenbaek, M.~Aspelmeyer, M.~Pfennigbauer, T.~Jennewein,
C.~Brukner, W.~R.
  Leeb, and A.~Zeilinger.
\newblock Proof-of-concept experiments for quantum physics in space.
\newblock {\em Proc. of SPIE}, 5161:252--268, 2003.

\bibitem{Gisin02a}
N.~Gisin, G.~Ribordy, W.~Tittel, and H.~Zbinden.
\newblock Quantum cryptography.
\newblock {\em Rev. Mod. Phys.}, 74(1):145--195, 2002.

\bibitem{Waks02}
E.~Waks, A.~Zeevi, and Y.~Yamamoto.
\newblock Security of quantum key distribution with entangled photons against
  individual attacks.
\newblock {\em Phys.\ Rev.\ A}, 65:52310, 2002.

\bibitem{Takesue05}
H.~Takesue, E.~Diamanti, T.~Honjo, C.~Langrock, M.~M. Fejer,
K.~Inoue, and
  Y.~Yamamoto.
\newblock Differential phase shift quantum key distribution experiment over 105
  km fibre.
\newblock {\em New Journal of Physics}, 7:232, 2005.

\bibitem{Valencia02}
A.~Valencia, G.~Scarcelli, and Y~Shih.
\newblock Distant clock synchronization using entangled photon pairs.
\newblock {\em Appl. Phys. Lett.}, 85:2655, 2004.

\bibitem{Naletto07}
G.~Naletto, C.~Barbieri, T.~Occhipinti, F.~Tamburini, S.~Billotta,
S.~Cocuzza,
  and D.~Dravins.
\newblock Very fast photon counting photometers for astronomical applications:
  from quanteye to aqueye.
\newblock In {\em Photon counting applications, Quantum Optics, and Quantum
  Cryptography. SPIE Proc. 6583, pp. 65830B-1/14}, 2007.

\bibitem{Carey01}
W.~Carey, D.~Isakeit, M.~Heppener, K.~Knott, and J.~Feustel-Bechl.
\newblock The international space station european users guide.
\newblock Technical report, Tech. Rep., European Space Agency, ISS User
  Information Centre (MSM-GAU), ESTEC, 2001.

\bibitem{Pfennigbauer05}
M.~Pfennigbauer, M.~Aspelmeyer, W.~Leeb, G.~Baister, T.~Dreischer,
  T.~Jennewein, G.~Neckamm, J.~Perdigues, H.~Weinfurter, and A.~Zeilinger.
\newblock Satellite-based quantum communication terminal employing
  state-of-the-art technology.
\newblock {\em J. Opt. Netw.}, 4(9):549--560, 2005.

\bibitem{aspelmeyer03a}
M.~Aspelmeyer, T.~Jennewein, M.~Pfennigbauer, W.~R. Leeb, and
A.~Zeilinger.
\newblock Long-distance quantum communication with entangled photons using
  satellites.
\newblock In {\em IEEE Journal of Selected Topics in Quantum Electronics
  1541-1551}, 2003.

\bibitem{Rarity02}
J.~G. Rarity, P.~R. Tapster, P.~M. Gorman, and P.~Knight.
\newblock Ground to satellite secure key exchange using quantum cryptography.
\newblock {\em New Journal of Physics}, 4:82, 2002.

\bibitem{Poppe08}
A.~Poppe, M.~Peev, and O.~Maurhart.
\newblock Outline of the secoqc quantum-key-distribution network in vienna.
\newblock {\em to appear in Int. J. Quant. Inf.}, 2008.

\bibitem{Dianeti08}
M.~Dianati and R.~Alléaume und M. Gagnaire~und X.~Shen.
\newblock Architecture and protocols of the future european quantum key
  distribution network.
\newblock {\em Security and Communication Networks}, 1:57--74, 2008.

\bibitem{Acin07}
A.~Acin, J.~I. Cirac, and M.~Lewenstein.
\newblock Entanglement percolation in quantum networks.
\newblock {\em Nature Physics}, 3:256--259, 2007.

\bibitem{Giggenbach07}
N.~Perlot, M.~Knapek, D.~Giggenbach, J.~Horwath, M.~Brechtelsbauer,
  Y.~Takayama, and T.~Jono.
\newblock Results of the optical downlink experiment {KIODO} from {OICETS}
  satellite to optical ground station oberpfaffenhofen ({OGS}-{OP}).
\newblock In {\em Conference on Laser Communication and Propagation, Proc. of
  SPIE 6457A}, 2007.

\bibitem{Toyoshima07b}
M.~Toyoshima, T.~Takahashi, K.~Suzuki, S.~Kimuraa, K.~Takizawa,
T.~Kuri,
  W.~Klaus, M.~Toyoda, H.~Kunimori, Y.~T.~Jono, Takayama, and K.~Arai.
\newblock Results from phase-1, phase-2 and phase-3 kirari optical
  communication demonstration experiments with the nict optical ground station
  (koden).
\newblock In {\em 24th International Communications Satellite Systems
  Conference of AIAA, AIAA-2007-3228, Korea}, 2007.

\bibitem{kurtsiefer02}
C.~Kurtsiefer, P.~Zarda, M.~Halder, H.~Weinfurter, P.~M. Gorman,
P.~R. Tapster,
  and J.~G. Rarity.
\newblock A step towards global key distribution.
\newblock {\em Nature}, 419:450, 2002.

\bibitem{hughes02}
R.~J. Hughes, J.~E. Nordholt, D.~Derkacs, and G.Peterson.
\newblock Practical free-space quantum key distribution over 10 km in daylight
  andat night.
\newblock {\em New Journal of Physics}, 4:43, 2002.

\bibitem{Aspel03c}
M.~Aspelmeyer, H.~B\"ohm, T.~Gyatso, T.~Jennewein, R.~Kaltenbaek,
  M.~Lindenthal, G.~Molina-Terriza, A.~Poppe, K.~Resch, M.~Taraba, R.~Ursin,
  P.~Walther, and A.~Zeilinger.
\newblock Long-distance free-space distribution of entangled photons.
\newblock {\em Science}, 301:621--623, 2003.

\bibitem{pan05a}
C.~Z. Peng, T.~Yang, X.~H. Bao, J.~Z, X.~M.~Jin andF. Y.~Feng,
J.~Yang, J.~Yin,
  Q.~Zhang, N.~Li, B.~L. Tian, and J.~W. Pan.
\newblock Experimental free-space distribution of entangled photon pairs over a
  noisy ground atmosphere of 13km.
\newblock {\em Phys. Rev. Lett.}, 94:150501, 2005.

\bibitem{Schmitt07}
T.~Schmitt-Manderbach, H.~Weier, M.~Fürst, R.~Ursin,
F.~Tiefenbacher,
  T.~Scheidl, J.~Perdigues, Z.~Sodnik, C.~Kurtsiefer, J.~G. Rarity,
  A.~Zeilinger, and H.~Weinfurter.
\newblock Experimental demonstration of free-space decoy-state quantum key
  distribution over 144 km.
\newblock {\em Phys. Rev. Lett.}, 98:010504, 2007.

\bibitem{ursin07}
R.~Ursin, F.~Tiefenbacher, T.~Schmitt-Manderbach, H.~Weier,
T.~Scheidl,
  M.~Lindenthal, B.~Blauensteiner, T.~Jennewein, J.~Perdigues, P.~Trojek,
  B.~Oemer, M.~Fuerst, M.~Meyenburg, J.~Rarity, Z.~Sodnik, C.~Barbieri,
  H.~Weinfurter, and A.~Zeilinger.
\newblock Entanglement-based quantrum communication over 144 km.
\newblock {\em Nature Physics}, 3:481 -- 486, 2007.

\bibitem{Villoresi08}
P.~Villoresi, T.~Jennewein, F.~Tamburini, M.~Aspelmeyer, C.~Bonato,
R.~Ursin,
  C.~Pernechele, V.~Luceri, G.~Bianco, A.~Zeilinger, and C.~Barbieri.
\newblock Experimental verification of the feasibility of a quantum channel
  between space and earth.
\newblock {\em New J. Phys.}, 10:033038, 2008.

\bibitem{Fedrizzi07}
A.~Fedrizzi, T.~Herbst, A.~Poppe, T.~Jennewein, and A.~Zeilinger.
\newblock A wavelength-tunable fiber-coupled source of narrowband entangled
  photons.
\newblock {\em Opt. Express}, 15(23):15377--15386, 2007.

\bibitem{Perdigues07}
J.~Perdigues, B.~Furch, C.~de~Matos, O.~Minster, L.~Cacciapuoti,
  M.~Pfennigbauer, M.~Aspelmeier, T.~Jennewein, R.~Ursin,
  T.~Schmitt-Manderbach, G.~Baister, J.~Rarity, W.~Leeb, C.~Barbieri,
  H.~Weinfurter, and A.~Zeilinger.
\newblock Quantum communication at {ESA}: Towards a space experiment on the
  {ISS}.
\newblock In {\em Coference Proceedings IAC2007 Hydarabath, India, accepted for
  publication in Acta Astronautica}, 2007.

\end{thebibliography}
\end{document}